# Constraint Minimum Vertex Cover in *K*-Partite Graph: Approximation Algorithm and Complexity Analysis

Kamanashis Biswas
Computer Science and Engineering Department
Daffodil International University
102, Shukrabad, Dhaka-1207
ananda@daffodilvarsity.edu.bd

S.A.M. Harun
Right Brain Solution
Flat# B4, House# 45, Road# 27
Banani, Dhaka
harun@rightbrainsolution.com

*Abstract* – Generally, a graph *G*, an independent set is a subset *S* of vertices in *G* such that no two vertices in *S* are adjacent (connected by an edge) and a *vertex cover* is a subset *S* of vertices such that each edge of *G* has at least one of its endpoints in *S*. Again, the *minimum vertex cover problem* is to find a vertex cover with the smallest number of vertices. Consider a *k*-partite graph $G = (V, E)$ with vertex *k*-partition $V = P_1 \cup P_2 \ldots \cup P_k$ and the *k* integers are $k_{p1}, k_{p2}, \ldots, k_{pk}$. And, we want to find out whether there is a minimum vertex cover in *G* with at most $k_{p1}$ vertices in $P_1$ and $k_{p2}$ vertices in $P_2$ and so on or not. This study shows that *the constrained minimum vertex cover problem in k-partite graph (MIN-CVCK) is NP-Complete* which is an important property of *k*-partite graph. Many combinatorial problems on general graphs are NP-complete, but when restricted to *k*-partite graph with at most *k* vertices then many of these problems can be solved in polynomial time. This paper also illustrates *an approximation algorithm for MIN-CVCK and analyzes its complexity*. In future work section, we specified a number of dimensions which may be interesting for the researchers such as developing algorithm for maximum matching and polynomial algorithm for constructing *k*-partite graph from general graph.

## I. INTRODUCTION

NP-Completeness theory is one of the most important developments of algorithm research since its introduction in the early 1970. Its importance arises from the fact that the results have meaning for all researchers who are developing computer algorithms, not only computer scientist but also for the electrical engineers, operation researchers etc. A wide variety of common encountered problems from mathematics, computer science and operations research are known to be NP-Complete and the collection of such problems is continuously rising almost everyday. Indeed, the NP-Complete problems are now so pervasive that it is important for anyone concentrated with the computational aspect of these fields to be familiar with the meaning and implementations of this concept. A number of works have already been done as well as going today. For example, Jianer Chen et al. have shown that the complexity of an algorithm for solving vertex cover problem is non deterministic polynomial [3]. Again, the complexity of algorithm of constrained minimum vertex cover in bipartite graph is also non deterministic polynomial which is proved by Jianer Chen & Iyad A. Kanj [2]. Similarly, H. Fernan & R. Niedermeier has proposed an efficient exact algorithm for constrained bipartite vertex cover is also non deterministic polynomial [4]. This paper shows that the minimum vertex cover in *k*-partite graph is NP-Complete, provides an approximation algorithm and analyzes its complexity which is polynomial time algorithm.

## II. PRELIMINARY

This section presents some basic terms and necessary elaborations which are important to go though the rest of the paper. Definitions that are not included in this section will be introduced as they are needed.

### A. Bipartite Graph

A *bipartite graph* is any graph whose vertices can be divided into two sets, such that there are no edges between vertices of the same set [8]. A graph can be proved bipartite if there does not exist any circuits of odd length. A set of vertices decomposed into two disjoint sets such that no two vertices within the same set are adjacent. A *bigraph* is a special case of a *k*-partite graph with *k* = 2.

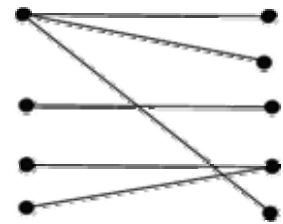

Figure 2.1: *Bipartite Graph*

### B. K-partite Graph

A *k-partite Graph* (i.e., a set of vertices decomposed into *k* disjoint sets such that no two vertices within the same set are adjacent) such that every pair of vertices in the *k* sets are adjacent [9]. If there are *p, q, . . . , r* vertices in the *k* sets, the complete *k*-partite graph is denoted $k_{p,q,\ldots,r}$.

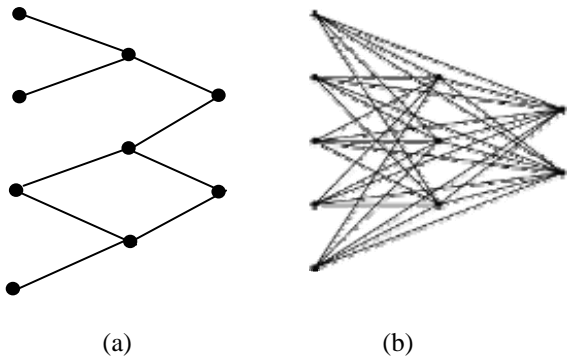

(a)　　　　　　　　　(b)

Figure 2.2: (a) *k-partite graph,* (b) *Complete k-partite graph*

## C. Vertex Cover

Let $S$ be a collection of subsets of a finite set $X$. The smallest subset $Y$ of $X$ that meets every member of $S$ is called the *vertex cover*, or hitting set. However, some authors call any such set a vertex cover, and then refer to the minimum vertex cover [6]. Finding the hitting set is an NP-Complete problem. Vertex covers, indicated with no fill vertices, are shown in the figure 2.3 for a number of graphs. In a complete *k*-partite graph, vertex cover contains vertices from at least **K-1** stages.

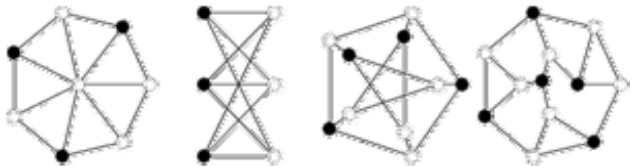

Figure 2.3: *Vertex Cover*

*1. Minimum Vertex Cover:* As a detailed example of an NP-Complete problem in Section III, we have described the VERTEXCOVER problem. Given a graph $G = (V, E)$, is there a vertex cover, i. e., a subset of nodes, that touches all edges in $E$ and contains not more than $k$ vertices, where $k$ is a given constant? Posed as a language, this problem becomes-

VERTEX-COVER = {$(G, K)$ | $G$ has vertex cover of at most $k$ vertices.}

Typically, this problem can be asked in another form: Instead of asking whether some vertex cover exists, the task is to find the smallest possible vertex cover: Given a graph $G = (V, E)$, find a vertex cover of $G$ having the smallest possible number of vertices. Typically, optimization problems like this one are even more difficult to decide than the related yes/no problems: VERTEXCOVER is NP-Complete, but MIN-VERTEX-COVER is NP-hard, i. e., it is not even in NP itself.

*2. Approximate Vertex Cover:* We know that finding the minimum vertex cover of a graph is NP-Complete. However, a very few simple procedures can efficiently find a vertex cover that at most twice as large as the optimal cover. Let us see a simple procedure [6].

VertexCover ($G = (V, E)$)
　While ($E \neq \emptyset$) do:
　　Select an arbitrary edge $(u, v) \leq E$
　　Add both $u$ and $v$ to the vertex cover
　　Delete all edges from $E$ that are
　　incident on either $u$ or $v$.

*3. Constraint Vertex Cover:* The *constrained vertex cover* of an undirected graph $G = (V, E)$ is a subset $V' \subseteq V$ where the number of vertex is less than or equal to $k$ [here $k$ is $k_{p1} + k_{p2} + \ldots + k_{pk}$]. That is, $V' \leq k$. We have to decide whether there is a minimum vertex cover in $G$ with at most $k_{p1}$ vertices in $P_1$ part and $k_{p2}$ vertices in $P_2$ and so on.

## D. Class P and NP

The class P is the type of problems that can be solved by polynomial time algorithm. For the problem of class P, polynomial time algorithm already exists. For example matrix multiplication algorithm, Prim's minimum spanning tree algorithm, graph traversal algorithm etc. are polynomial time algorithm. On the other hand, the name NP stands for nondeterministic polynomial. The class NP is the set of problems that can be solved by nondeterministic algorithm in polynomial time or the set of problems whose solution can be verified by a polynomial time algorithm [5]. No deterministic polynomial time algorithm exists for the problems of NP class.

## E. Properties of NP-Complete Problem

Let $L_1$ and $L_2$ be two problems. $L_1$ reduces to $L_2$ (also written $L_1 \leq_p L_2$) if and only if there is a way to solve $L_1$ by a deterministic polynomial time using a deterministic algorithm that solves $L_2$ in polynomial time [7]. We can now define the set of NP-Complete problems, which are the hardest problems in NP in the following ways. A problem $L$ is NP-Complete if-
　1. $L \in$ NP, and
　2. $L^1 \leq_p L$ for $L^1 \in$ NPC.
That is, more precisely we can say a problem in NP-Complete if and only if-
　1. The problem is in NP and
　2. The problem is polynomial reducible from another problem that is already in NP-Complete.
If a problem $L$ satisfies property 2, but not necessarily property 1, then we say that $L$ is NP-hard.

## III. VERTEX COVER AND CLIQUE PROBLEM

The vertex cover of an undirected graph $G = (V, E)$ is a subset $V' \subseteq V$ such that if $(u, v) \in E$, then $u \in V'$ or $v \in V'$ (or both). More precisely, it is the optimization problem of finding a *vertex cover of minimum size in a graph* that is finding a minimum number of vertices that "covers" all edges. The following figure illustrates minimum vertex cover of the graph $G$.

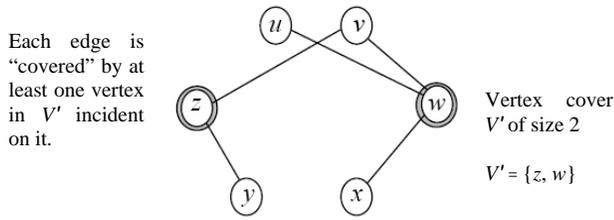

Each edge is "covered" by at least one vertex in V' incident on it.

Vertex cover V' of size 2

V' = {z, w}

Figure 3.1: *Minimum vertex cover of graph G with size V'*

A *clique* in an undirected graph $G = (V, E)$ is a subset $V' \subseteq V$ of vertices, each pair of which is connected by an edge in $E$. Similar to vertex cover problem, the *clique problem* is also the optimization problem of finding a clique of maximum size in a graph. The practical usage of clique problem is in synthesis and optimization of digital systems to model certain resource allocation constraints etc.

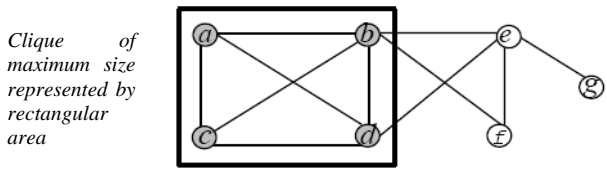

*Clique of maximum size represented by rectangular area*

Figure 3.2: *Clique Problem*

## IV. MAIN THEOREM

It is already proved that the MIN-CVCB (Constrained minimum vertex cover in bipartite graph) problem is in NP. Jianer Chen and Iyad A. Kanj have proved the theorem in "Constrained minimum vertex cover in bipartite graphs: complexity and parameterized algorithms" in 2003 [2]. G. Bai and H. Fernau show that exact algorithms could perform much better than theoretical assumption [1]. In this section, the main theorem of this research is described which shows that vertex cover in *k*-partite graph is NP-Complete.

**Theorem:** *The minimum constrained vertex cover problem is NP-Complete in k-partite graph.*

**Proof:** We show that VERTEX-COVER $\in$ NP. Suppose, we are given a graph $G = (V, E)$ with vertex *k*-partition $V = P_1 \cup P_2 \ldots \cup P_k$ and the integers $k_{p1}, k_{p2}, \ldots, k_{pk}$ where $k = k_{p1} + k_{p2} + \cdots + k_{pk}$. The certificate we choose if the vertex cover $V' \subseteq V$ itself. The verification algorithm affirms that $|V'| = k$, and then it checks, for each edge $(u, v) \in E$, whether $u \in V'$ or $v \in V'$. This verification can be performed straightforwardly in polynomial time.

We can prove that the Vertex-cover problem is NP-hard by showing that CLIQUE $\leq_p$ VERTEX-COVER. This reduction is based on the notion of the "complement" of a graph. Given an *k*-partite graph $G = (V, E)$, the complement of $G$ is defined as $\overline{G} = (V, \overline{E})$ where $\overline{E} = \{(u, v) : (u, v) \notin E\}$. In other words, $\overline{G}$ is the graph containing exactly those edges that are not in $G$.

The figure 4.1 shows a graph and its complement and illustrates the reduction from CLIQUE to VERTEX-COVER.

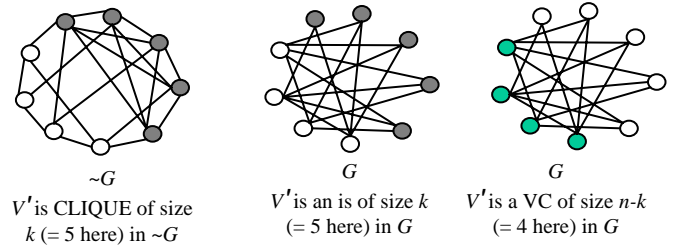

~G
V' is CLIQUE of size $k$ (= 5 here) in ~G

G
V' is an is of size $k$ (= 5 here) in G

G
V' is a VC of size $n-k$ (= 4 here) in G

Figure 4.1: *3 Easy reductions*

The reduction algorithm takes as input an instance $(G, k)$ of the clique problem. It computes the complement $\overline{G}$, which is easily double in polynomial time. The output of the reduction algorithm is the instance $(\overline{G}, |V| - K)$ of the vertex-cover problem. To complete the proof, we show that this transformation is indeed a reduction: the *k*-partite graph has a clique of size $k$ if and only if the graph $\overline{G}$ has a vertex cover of size $|V| - k$ as shown in the figure 4.1.

- *Instance <G, k> of CLIQUE*

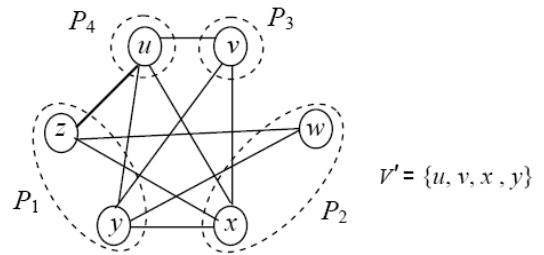

V' = {u, v, x, y}

Figure 4.2: *4-partite graph*

Here, the above graph $G$ is a 4-partite graph. Suppose that $G$ has a clique $V' \subseteq V$ with size $k = |V'|$. The subsets produced in the previous graph are as follows:

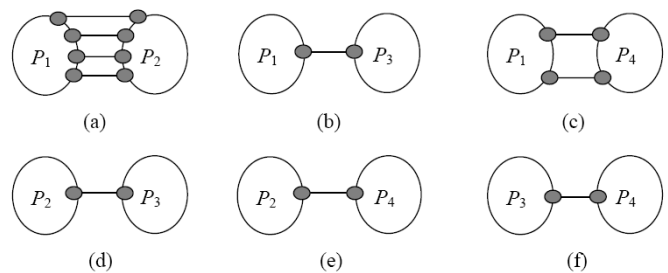

(a)   (b)   (c)

(d)   (e)   (f)

Figure 4.3: Subsets of graphs produced form 4-partite graph

- *Instance <$\overline{G}$, |V| - k > of VERTEX-COVER*

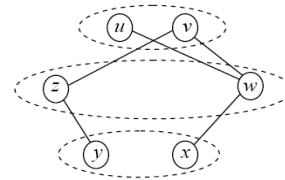

Figure 4.4: 3-partite graph

So $V - V'$ is the vertex cover in $\overline{G}$ with size $|V| - k$.

Let $(a, b)$ be any edge in $\overline{E}$, then $(a, b) \notin E$, which implies that at least one of $a$ or $b$ does not belong to $V'$. Since every pair of vertices in $V'$ is connected by an edge of $E$. Equivalently, at least one of $a$ or $b$ is in $V - V'$, which means that the edge $(a, b)$ is covered by $V - V'$. Since $(a, b)$ was chosen arbitrarily from $\overline{E}$, every edge of $\overline{E}$ is covered by a vertex in $V - V'$. Hence the set $V - V'$ has size $|V| - k$, from a vertex cover for $\overline{G}$.

Conversely, suppose that $\overline{G}$ has a vertex cover $V' \subseteq V$, where $|V'| = |V| - k$, then for all $a, b \in V$, if $(a, b) \in \overline{E}$ then $a \in V'$ or $b \in V'$ or both. If $a \in V'$ and $b \notin V'$ then $(a, b) \in E$, $V - V'$ is in a clique and it has size $|V| - |V'| = k$.

## V. APPROXIMATION ALGORITHM

At present, all known algorithms for NP-Complete problems require time which is exponential. It is unknown whether there are any faster algorithms. Therefore, in order to solve an NP-Complete problem for any non-trivial problem size, one of the following approaches is used according to [6]:

- Approximation: An algorithm which quickly finds a suboptimal solution which is within a certain (known) range of the optimal one. Not all NP-Complete problems have good approximation algorithms, and for some problems finding a good approximation algorithm is enough to solve the problem itself.

- Probabilistic: An algorithm which probably yields good average runtime behavior for a given distribution of the problem instances—ideally, one that assigns low probability to "hard" inputs.

- Special cases: An algorithm which is probably fast if the problem instances belong to a certain special case.

- Heuristic: An algorithm which works "reasonably well" on many cases, but for which there is no proof that it is always fast.

Approximation algorithms return solutions with a guarantee attached, namely that the optimal solution can never be much better than this given solution. Thus we can never go too far wrong in using an approximation algorithm. No matter what our input instance is and how lucky we are, we are doomed to do all right. Further, approximation algorithms realizing provably good bounds often are conceptually simple, very fast, and easy to program.

### A. Algorithm for MIN-CVCK problem

In this section, we described our proposed algorithm for minimum constrained vertex cover in $k$-partite graph. It is an approximation algorithm for MIN-CVCK problem. The procedure is described in the next column.

### Proposed algorithm for CVCK problem

*Procedure MIN-CVCK (n, G, U, [Count], K)*

[ // $n$ is the number of partition, $n \geq 2$
// $G$ is a given Graph
// $U$ is the list of vertex in each partition which is size of $n$
// Count is an array contains how many vertices in each partition
// $K$ is the array which indicates we can take at most $K[i]$ vertices from the $i$-th partition]

Integer $a[]$, $b[]$, part[], tmpU[]
Struct EdgeList[]

[ // $a$ is a flag array which track whether a vertex is selected or not selected or not used
// $a$ is initialized with not used
// $b$ is an integer array which contains how many vertices are already used by each partition.
// $b$ is initialized with 0
// part is an array indicates the partition in which a vertex lies
// EdgeList is an array of structure containing edge]

$G' = G$ // Compute part array from $U$

while (True)
{
 tmpU = Extract_max ($G'$)
// find a vertex $u$ from $G'$ with maximum degree ($\geq 1$) $u \in G[v]$
If tmpU = NULL Then Break
Else If $b[$ part $[$ tmpU $]$ $] + 1 > K[$ part $[$ tmpU $]$ $]$
// here part[tmpU] is the partition where tmpU lies
$a$[tmpU] = not selected
Else $a$[tmpU] = selected
$b[$ part $[$ tmpU $]$ $] = b[$ part $[$ tmpU $]$ $] + 1$
EdgeList = NULL
remove all the coincident edge of tmpU from $G'$ and add those edges to EdgeList
If Make_decision ($G'$) = False
$a$[tmpU] = not selected
$b[$ part $[$ tmpU $]$ $] = b[$ part $[$ tmpU $]$ $] - 1$
add the edges in EdgeList to $G'$
 End If
} // End_While
// End_Procedure_MIN-CVCK

*Procedure node_type Extract _max (G)*
{
Max = 0, MaxDegVertex = NULL
for each vertex in $V[G]$
if $a[u]$ = not used and degree[$u$] > Max Then
MaxDegVertex = $u$
Max = degree[$u$]
return MaxDegVertex
 } // End_ procedure_Extract_max

*Procedure boolean Make_decision(G)*
{
 Set $S$ = NULL // *S is a set of vertices*

For $i$ = 1 to Number of partition in $G$

Select none or [1, $k$ [$P$ [ ord [ $i$ ] ] ] $-$ $b$ [ $P$ [ ord [ $i$ ] ] ] ] vertices which are not used from $i$-th partition and add to S where every vertex coincident on at least one non-visited edge and more non-visited edges than those are not selected. Mark all edges as visited concerned with selected vertex.

[ // ord is an array containing partition number such that $k[$ $P$[ord[ $i$ ] ] ] $-$ $b[$ $P$[ord[ $i$ ] ] ] $\geq k$ [ $P$ [ ord [ $i$ +1 ] ] ] $-$ $b$ [ $P$ [ ord [ $i$ + 1 ] ] ] for all $i$ < $k$ ]

If there exist at least one edge in *G* not coincident on vertex $u \in S$
then return False
End If
Return True
} *// End_Procedure_Make_decision*

### B. Complexity Analysis

Here we will define the complexity of our proposed CVCK algorithm. Let us given a graph *G* (*V*, *E*) with k partition where $|V| = n$.

Now we get the complexity for average case,

$(n-1) \{ \log n + (n-1) + (n-1) + k + k(n-1) \}$
$\Rightarrow (n-1) \log n + 2(n-1)^2 + (n-1)k + k(n-1)^2$
$\Rightarrow (n-1)^2 (2+k) + (n-1)(k+\log n)$ ........ (i)

Complexity for best case-
When $k = 2$
then the complexity we get from equation (i),
$(n-1)^2 (2+2) + (n-1)(2+\log n)$
$\Rightarrow 4(n-1)^2 + (n-1) \log(n+4)$
$\Rightarrow O(n^2) + O(n \log n)$
$\Rightarrow O(n^2)$

Complexity for worst case-
When $k = n$
then the complexity we get from equation (i),
$(n-1)^2 (2+n) + (n-1)(2+\log n)$
$\Rightarrow O(n^3) + O(n^2) + O(n \log n)$
$\Rightarrow O(n^3)$

Hence, we have showed that the time complexity for the above CVCK approximation algorithm is $O(n^2)$. The following table summarizes some known results of vertex cover problems.

**Table 5.1:** Complexity of some vertex cover problems

| Problem Domain | Time | Reference |
|---|---|---|
| Vertex Cover Problem | $O(kn + 1.2852^k)$ | Jianer Chen, Weijia Jia & Iyad A. Kanj [3] |
| Constrained minimum vertex cover in bipartite graph | $O(1.26^{k_u+k_l}) + (k_u+k_l)|G|)$ | Jianer Chen & Iyad A. Kanj [2] |
| An efficient exact algorithm for constrained bipartite vertex cover | $O(1.40^k + kn)$ | H. Fernan & R. Niedermeier [4] |
| Constrained minimum vertex cover in *k*-partite graph | $O(n^2)$ | Ours |

## VI. CONCLUSION

Most theoretical computer scientist believes that the NP-complete problems are intractable. The reason is that if any single NP-complete problem can be solved in polynomial time, then every NP-complete problem has a polynomial-time algorithm. In this research, we show that the minimum vertex covering for *k*-partite graph is NP-complete. There are some limitations as our approximation algorithm is efficient for 80% graph. These are: i) if the graph can be drawn as a tree then our algorithm will give minimum + 1 solution for it, ii) there may be no solution or output for a very complex graph. Now, some of the open problems are as follows:

1. What is the complexity of constrained minimum vertex cover in *k*-partite graph?
2. Is it possible to minimize the complexity of approximation algorithm of this problem into $O(n \log n)$ or less from $O(n^2)$ ?

### A. Future Work

Is it possible to develop a perfect algorithm for maximum matching in *k*-partite graph? If it becomes possible then it will be easier to solve this type of NP-Complete problem. Is it possible to prove vertex cover in *k*-partite graph with node capacity (i.e. each node has its own costs) is NP-Complete? Is there any polynomial algorithm for construct *k*-partite graph from general graph?

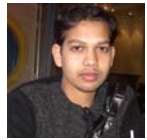
**Kamanashis Biswas**, born in 1982, post graduated from Blekinge Institute of Technology, Sweden in 2007. His field of specialization is on Security Engineering. At present, he is working as a Lecturer in Daffodil International University, Dhaka, Bangladesh.

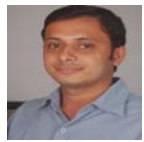
**S.A.M. Harun,** is graduated from International Islamic University Chittagong. He is a programmer and ACM problem setter. Now he is working as a project manager in a software company. His major area of interest is developing efficient algorithm.